\documentclass{article}

\usepackage{arxiv}
\usepackage[utf8]{inputenc}
\usepackage[T1]{fontenc}
\usepackage{amsmath}
\usepackage{graphicx}
\usepackage{rotating}
\usepackage[authoryear,round]{natbib}
\usepackage{microtype}
\usepackage{url}
\usepackage{hyperref}

\renewcommand{\shorttitle}{Long-window 4DVar for reanalysis}
\renewcommand{\headeright}{A Preprint}
\renewcommand{\undertitle}{A Preprint}

\title{Long-window 4DVar for reanalysis using a differentiable weather model}

\author{
  Gregory J. Hakim\textsuperscript{1},
  Jeffrey S. Whitaker\textsuperscript{2,3},
  Bo Huang\textsuperscript{4,5}, and
  Sergey Frolov\textsuperscript{5}\\[0.75em]
  \begin{minipage}{0.95\textwidth}
    \centering\normalfont\small
    \textsuperscript{1}Department of Atmospheric \& Climate Science, University of Washington, Seattle, Washington, USA\\
    \textsuperscript{2}UCAR Cooperative Programs for the Advancement of Earth System Science (CPAESS)\\
    \textsuperscript{3}NOAA/NWS Office of Modeling and Development, College Park, Maryland, USA\\
    \textsuperscript{4}Cooperative Institute for Research in Environmental Sciences, University of Colorado Boulder, Boulder, Colorado, USA\\
    \textsuperscript{5}NOAA Physical Sciences Laboratory, Boulder, Colorado, USA\\[0.5em]
    Corresponding author: Gregory J. Hakim (\href{mailto:ghakim@uw.edu}{ghakim@uw.edu})
  \end{minipage}
}

\date{}

\begin{document}

\maketitle

\begin{abstract}
Atmospheric reanalyses combine observations with model forecasts using complex data assimilation systems. We test whether a differentiable weather model permits a simpler and more accurate method based on a long-window four-dimensional variational data assimilation (4D-Var) formulation that omits the conventional background-error term. The method uses automatic differentiation to find optimal NeuralGCM initial conditions that minimize the misfit to real surface-pressure observations distributed across overlapping windows of two to seven days, assuming no model error. Cycling at 6-hour intervals for three months beginning 1 January 2015 yields a stable reanalysis with smaller error relative to ERA5 in 500-hPa geopotential height than the Twentieth Century Reanalysis version 3 (20CRv3), which uses an ensemble Kalman filter to assimilate the same observations. Every window produces smaller errors than 20CRv3, with analysis error for the four-day window approximately 55\% smaller than for 20CRv3. At the end of the four-day window, which does not benefit from future observations, error remains approximately 38\% smaller than 20CRv3. Analyses degrade slightly beyond four days, which we attribute to the increasing importance of model error.
\end{abstract}

\section*{Plain Language Summary}
Reconstructing gridded atmospheric structure from noisy observations and forecast models is a computationally demanding problem. Standard methods require complex estimates relating errors across locations and atmospheric variables. We show that a hybrid physics--machine-learning weather model can drastically simplify the calculation, while also improving accuracy. Neglecting the complex error estimates, and considering only observed surface pressure observations from land stations, ships, and buoys over 2--7 day windows for three months in 2015 yields stable analyses. Moreover, compared with NOAA's Twentieth Century Reanalysis version 3, which used the same observations, errors relative to the ERA5 dataset are about 55\% smaller in the analysis, and 38\% smaller at the end of the window, which does not benefit from later observations. Performance degrades for windows longer than four days due to model error, which is not accounted for in this study.

%
%

%


%
%
%
%

\section{Introduction \label{sec:introduction}}

Atmospheric reanalyses combine heterogeneous, error-prone observations with a forecast model and a data assimilation system to produce spatially and temporally complete estimates of the evolving atmosphere \citep[e.g.,][]{Parker2016,hersbach2020era5}. They underpin studies of weather and climate variability and provide training data for recently developed machine-learning weather models \citep[e.g.,][]{lam2022,kurth2023,bi2023,Kochkov_2024}. Because global data assimilation involves state vectors and observing systems with tens of millions of degrees of freedom, practical methods require approximations that balance statistical fidelity with computational cost \citep[e.g.,][]{Carrassi2018}. Efficient, differentiable weather models create new opportunities to revisit these trade-offs. Building on the use of differentiable weather models for computing shadowing trajectories \cite{heatwave2024,vonich2026}, we propose a four-dimensional variational data assimilation (4D-Var) approach that uses the efficiency and differentiability of NeuralGCM \cite{Kochkov_2024} to significantly simplify the assimilation algorithm. We evaluate the new algorithm in the Twentieth Century Reanalysis (20CR) framework, which assimilates only surface-pressure observations while prescribing sea-surface temperature and sea-ice boundary conditions \cite{Compo2011,Slivinski_2019}.

Data assimilation methods can be viewed broadly as filters or smoothers \cite{Carrassi2018}. A filter estimates the state sequentially by combining observations available at an analysis time with a forecast from the preceding analysis; the forecast carries the memory of information from earlier observations forward in time under the model dynamics. A smoother instead estimates a trajectory over a time window, allowing observations later in the window to inform states at earlier times. 4D-Var is a variational smoother that seeks an initial state whose model trajectory fits observations distributed through the window, usually constrained to remain consistent with a prior, or ``background,'' state \cite{LeDimetTalagrand1986,CourtierEtAl1994}.

In conventional atmospheric 4D-Var, the background-error covariance matrix describes prior uncertainty and its relationships among variables and locations. It controls the relative influence and spatial structure of analysis increments, and spreads information from observations to unobserved variables and regions. The background covariance can be viewed as a constraint on the analysis. For example, it promotes dynamical balances, such as gradient wind and hydrostatic balance. More generally, it limits the large space of potential analyses given the observations to a smaller space constrained by relationships encoded in the covariance. For filters and 4D-Var systems having short windows (e.g. typically 12h), this constraint is essential and considerable algorithmic complexity and computational expense is devoted to carefully defining it \citep[e.g.,][]{Bannister2008}.

A radical aspect of our proposed algorithm involves neglecting the background error covariance. This is motivated by a long-window approach, where the contribution of the background to the total error becomes vanishingly small in the long-time limit \cite{PiresEtAl1996,Swanson1998,FisherAuvinen2012}. Fitting observations over increasingly long time intervals shrinks the distribution of plausible states at the start of the window that can lead to those observations in the future. Effectively, the dynamics of the model over the window constrain the error at the start of the window to fit future observations. Since model error is unavoidable, it imposes a practical limit to the window length in the absence of measures to account for it \citep[e.g.,][]{HowesEtAl2017,farchi2021}; lacking such measures, we make the ``strong constraint'' assumption. Thus, we test the hypothesis that, even with sparse surface-pressure observations, dynamical constraints from the model can regularize the analysis without an explicit background-error covariance.

The proposed algorithm also departs from the traditional incremental form of 4DVAR as used by the European Centre for Medium Range Weather Forecasts (ECMWF) \cite{CourtierEtAl1994}. Specifically, we use automatic differentiation to compute the gradient of the cost function that includes nonlinear forecasts with the NeuralGCM model \cite{heatwave2024,Bano-Medina2025,SolvikEtAl2025,tian2026}.  Because of this non-linearity the resulting shape of the optimization surface is no longer quadratic \cite{PiresEtAl1996} and we can not use conjugate gradients to search for the minimum. Instead, we use the Adam optimizer to follow the gradient of the nonlinear cost function to find the optimal initial condition. 
Each iteration consists of a forward forecast to evaluate the observation mismatch and a reverse pass to update the initial condition. This approach avoids the need to develop and maintain the tangent linear and adjoint codes, and avoids linear error growth during the forward propagation of the analysis correction with the tangent linear model. The distinctive combination evaluated here is thus long-window, strong-constraint 4D-Var without a conventional background term, using a global differentiable hybrid weather model and real surface-pressure observations.

The remainder of the paper describes the long-window 4D-Var algorithm in section \ref{sec:method}, presents results in section \ref{sec:results}, and summarizes the conclusions in section \ref{sec:conclusions}.

\section{Method \label{sec:method}}

Unlike traditional 4D-Var algorithms, we neglect the background term, which results in a much simpler problem. The cost function we seek to minimize consists of two terms, one minimizing the misfit to observations, and the other penalizing fast decaying noise in the initial condition:

\begin{equation}
  J(\mathbf{x}_0, \dots, \mathbf{x}_K)\, = \,
\sum_{k=0}^{K} (\mathbf{y}_k -
\mathcal{H}(\mathbf{x}_k))^{\rm T} \mathbf{R}_k^{-1} (\mathbf{y}_k -
\mathcal{H}(\mathbf{x}_k)) \, + \alpha_tN_{ob}\iiint_{V}\left(\frac{\partial \mathbf{x}_0}{\partial t}\right)^2 \cos\phi \,d\lambda \,d\phi \,d\sigma. 
\label{eqn:loss}
\end{equation}
\noindent Here $K$ is the number of observation times in the window, $\mathbf{x}_k$ is the model state at observation time $k$, and $N_{ob}$ is the total number of observations in the window. The first term on the right side, $J_o$, is evaluated at regularly spaced observation times.  The second term, $J_t$,  penalizes large tendencies in the first model time step, which helps prevent the accumulation of rapidly decaying error during minimization. Such errors are not otherwise captured in the loss as they are small relative to larger errors at the end of the window. It is implemented on the geopotential field by averaging the initial tendency variance over the global domain. The amplitude of $J_t$ ($\alpha_t$) is scaled by the total number of observations in the window, so its amplitude relative to $J_o$ does not change as the window length changes. Although formally valid in the $t \rightarrow\infty$ limit, neglecting the background, $J_b$, term in (\ref{eqn:loss}) that appears in the traditional formulation is an approximation for finite $t$ that we test with our experiments.

The state evolves under the action of a perfect model, NeuralGCM (2.8$^{\circ}$ deterministic version), from one observation time to the next
\[
\mathbf{x}_k = \mathcal{M}(\mathbf{x}_{k-1}).
\]
\noindent Sea-ice and sea-surface-temperature boundary conditions are taken from ECMWF Reanalysis v5 \citep[ERA5;][]{hersbach2020era5}. Time is defined for the observations by
\[
t = \Delta t \, k, \qquad k = 0, 1, \dots, K
\]
\noindent Two additional time scales are the time between analyses (the ``cycling interval''), and the length of the window size for the minimization ($K \Delta t$). We fix the cycling interval at 6 hours and the observation interval $\Delta t$ at 3 hours for all experiments.  The window size $K \Delta t$ is varied from 2 to 7 days.

Minimization of (\ref{eqn:loss}) is performed using JAX for automatic differentiation, and Optax for managing gradient descent using the AdamW optimizer \cite{loshchilov2019adamw}. The algorithm consists of a nonlinear forward pass to compute $J$ in the decoded (physical) space, followed by a backward gradient pass to update the encoded-space version of $\mathbf{x}_0$. After a fixed number of epochs, the final $\mathbf{x}_0$ is taken as the analysis. A forecast on the cycling interval advances the state to the next analysis time, where the minimization begins anew. This process is repeated 360 times starting 00 UTC 1 January 2015, which is initialized with a 48h forecast begun from ERA5 \cite{hersbach2020era5} on 00UTC 30 December 2014. Using Python/JAX \cite{jax2018github} in both the model and the data assimilation algorithm drastically simplifies the code, since nearly all of the complexity is encapsulated in the JAX/XLA libraries for auto-differentiation, JIT compilation and optimization.  It also obviates the need for using a tangent-linear model in the forward pass, allowing for the use of longer windows without violating linearity assumptions.

Note that the gradients of all moist variables (vapor, cloud water, and cloud ice) are set to zero in these experiments.  Including them in the minimization simply requires rescaling the moist variables so that the gradient magnitudes are of the same order as the dynamical variables.  The fact that the moist variable gradients were not included allows us to evaluate the stability of the moisture fields as a more stringent test of the algorithm, since these fields are only constrained by the action of the model through the forecast step from the analysis to the start of the next window (while the rest of the control space is constrained by the action of the model during minimization).

In the experiments described below, we vary the window size from 2 to 7 days, and use surface pressure observations from ships, buoys and land stations extracted from the operational NOAA archive. The observation interval is 3 hours, and only observations within 30 minutes of the nominal observation time are used, resulting in between 6,000 and 12,000 observations each observation time (see Figure S1 in Supporting Information for the spatial distribution of observations at 01 January 2015 at 12UTC).  We assume that observation errors are uncorrelated, and set the observation-error standard deviation to 1.6 hPa for marine and 1.0 hPa for land observations. The forward operator for surface pressure is computed by linearly interpolating/extrapolating the logarithm of pressure linearly in geopotential height to the station elevation, using the pressure-level geopotential height field in the decoded model state.  50 epochs are used for the optimization, and the AdamW learning rate and weight decay parameters are set to 0.0001 (the learning rate is ramped up linearly during the first 10 epochs).  Observations are rejected if the difference between the station elevation and ERA5 orography interpolated to the observation location is greater than 500 meters, or the observation innovation is greater than 4 times the observation error standard deviation.  Observation error standard deviations are inflated from the nominal value by 0.003 hPa for every meter difference between that station elevation and the ERA5 orography at the observation location.  Wall time per epoch on an NVIDIA H100 GPU with 94Gb of memory is roughly 7 seconds for the 7 day window, 4-5 seconds for the 4 day window, and 2 seconds for the 2 day window experiments, i.e., scaling is nearly linear in window length.

At each cycle, the analysis interval is advanced by six hours, regardless of the length of the window. This results in an overlap between successive assimilation windows.  As pointed out by \citet{FisherAuvinen2012}, there is no statistical inconsistency in taking the starting point of the minimization from the preceding analysis, even though that starting point is informed by most of the observations in the new assimilation window.  This is because there is no constraint for the optimized analysis to stay close to the starting point, since there is no background term in the loss function (\ref{eqn:loss}).  

\section{Results \label{sec:results}}

\begin{figure}
\includegraphics[width=0.8\textheight]{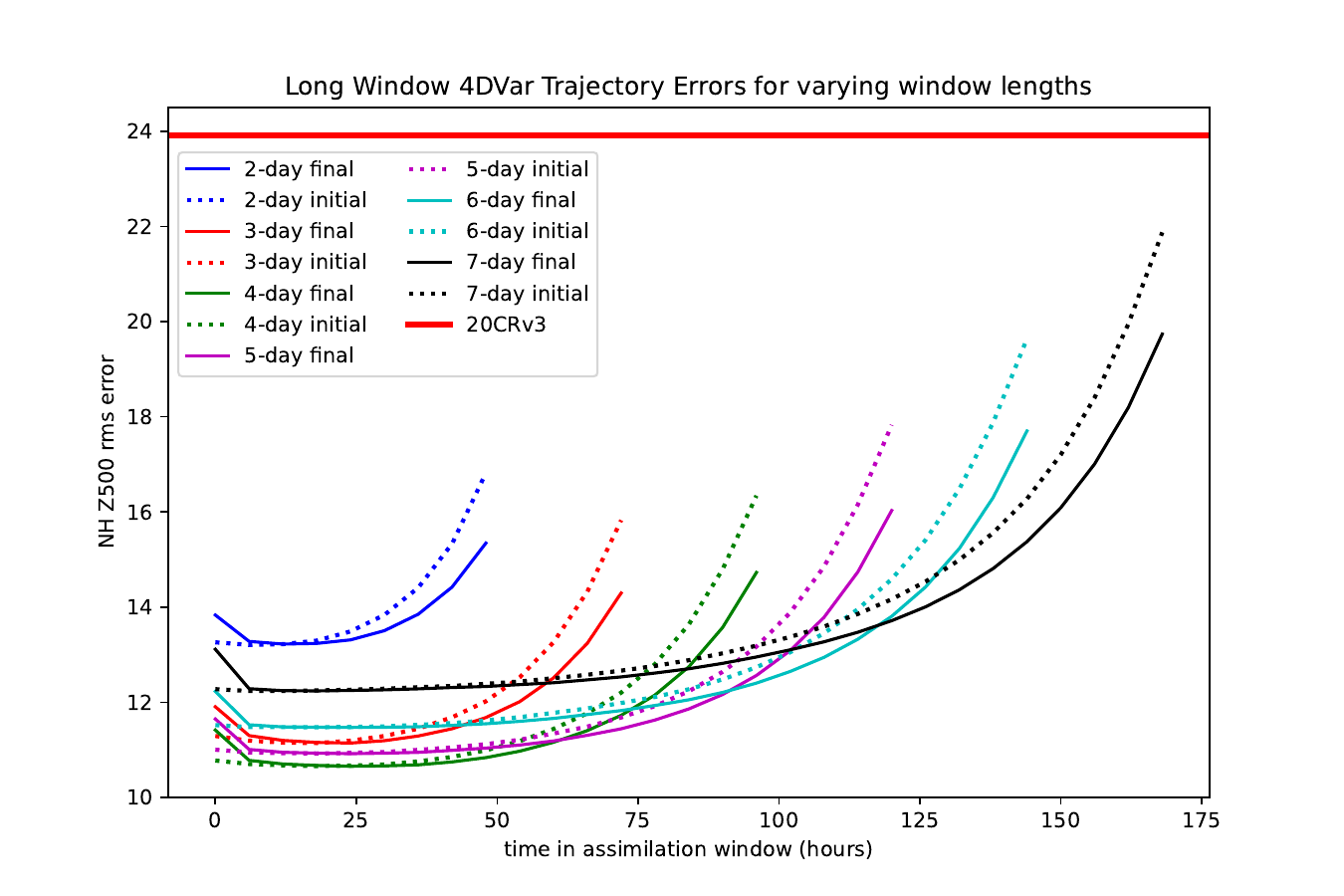}
\caption{Northern Hemisphere (poleward of $20^{\circ}$ N) 500 hPa geopotential height root mean squared error (relative to ERA5) over the assimilation window for long-window 4DVar experiments with varying window lengths from 2-7 days. The solid curves show the analysis error for the optimized trajectory (epoch 50), the dashed curves show the error for the initial trajectory (epoch 1). All calculations are performed on the 2.8$^{\circ}$ model grid, and are averaged over the 362 analyses between January 1 2015 at 00UTC and April 1 2015 at 06UTC.  The thick red horizontal line shows the 500 hPa geopotential height root mean squared error for 20CRv3 averaged over the same period as a reference.}
\label{z500err_window_err}
\end{figure}

Figure \ref{z500err_window_err} shows that, relative to ERA5, the 4DVar analysis global-mean root-mean squared 500 hPa geopotential height (Z500) errors for all window lengths (at all times in the window) are less than the 20th Century Reanalysis version 3 (20CRv3), with the minimum error occurring at 24-h into the 4-day window (10.65 meters versus 23.92 meters for 20CRv3, a 55\% reduction). These errors are averaged over 3 months of cycling for the initial (epoch 1) and final (epoch 50) trajectories for assimilation windows varying from 2 to 7 days in length.  As a reference, the time-averaged Z500 error for the same time period for 20CRv3 \cite{Slivinski_2019}, which assimilated the same observations, is shown by the red horizontal line.  Since 20CRv3 was produced with an ensemble Kalman filter that is only informed by observations prior to the nominal analysis time, it is perhaps fairer to compare 20CRv3 errors to the long-window 4DVar errors at the end of the window. The minimum error at the end of the 4-day window is 14.7 meters, still 38\% less than 20CRv3. This is despite the fact that the forecast model used in the long-window assimilation algorithm is much lower resolution (2.8$^{\circ}$ here versus approximately 0.75$^{\circ}$ for 20CRv3).  As the window length increases beyond 4 days, the 4DVar errors increase. One possibility is that this increase in error is due to the effects of model error, which is unaccounted for in the strong-constraint formulation.  Another possibility is that as windows become longer, the minimization overfits errors at the end of the window, which are relatively large, at the expense of errors earlier in the window. This motivates progressively expanding windows \cite{PiresEtAl1996,vonich2026}, which we have not yet tested.

The Z500 errors for all window lengths decrease sharply at the very beginning of the window. We interpret this as an artifact of not having a background term to penalize rapidly decaying features that improve the fit to observations early in the window, without affecting the fit to observations later in the window.  The amplitude of these rapidly decaying modes is penalized imperfectly by the $J_t$ term in the loss.  We have used an amplitude $\alpha_t=20$ for $J_t$ term, which is roughly the same amplitude as $J_{o}$ for a single observation time. This greatly reduces the amplitude of these rapidly decaying features, but does not eliminate them completely.  The sensitivity of the results to the amplitude of $J_t$ can be seen in Figure S2 in the Supporting Information section.  

The difference between the first and last epoch of the minimization (the dotted and solid curves in Figure \ref{z500err_window_err}) is largest at the end of the window, reflecting the fact that most of the improvement during the optimization derives from improving the fit of the trajectory to observations toward the end of the window, where errors have grown the most, and new observational information is being included. 

\begin{figure}
\includegraphics[width=0.8\textheight]{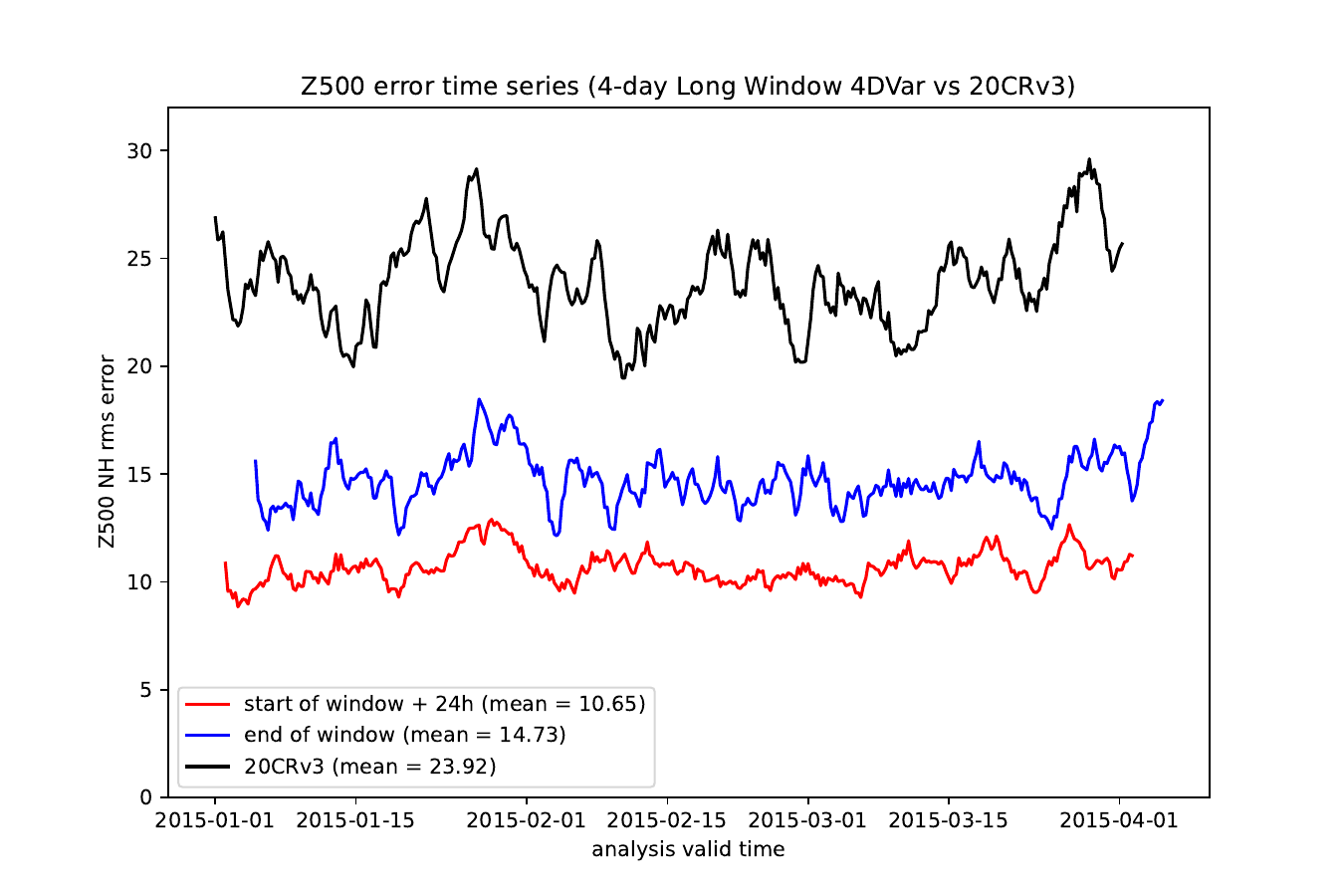}
\caption{Time series of Northern Hemisphere (poleward of $20^{\circ}$ N) 500 hPa geopotential height root mean squared error (relative to ERA5) for 4-day 4DVar with NeuralGCM 24-h into the window (red) and at the end of the window (blue), compared to 20CRv3 (black). Calculations are performed on the 2.8$^{\circ}$ model grid.}
\label{z500err_ts}
\end{figure}

Figure \ref{z500err_ts} shows time series of Z500 error at 24-h into the window and at the end of the window for 4-day 4DVar, compared to 20CRv3.  The 24-h result reflects the best smoothed state (informed by observations from all times in the window) and the end-of-window result reflects the filtered state (not informed by future observations).   These time series show that the cycling long-window 4DVar approach with overlapping windows is stable across the 3-month period, and that the improvement over 20CRv3 is high statistically significant (since there are no times when the 20CRv3 analysis is more accurate than 4DVar).  The error of the smoothed state estimate (the red curve, which includes observations both before and after the nominal analysis time) is about 27\% less than the filtered state estimate (the blue curve, which only includes observational information prior to the nominal analysis time).
Forecasts from the long-window 4DVar analyses relative to the end of the assimilation window show similar error growth to forecasts initialized from ERA5, with about a 2--3-day lag, even though they are only informed by surface pressure observations (Figure S3). 

\begin{figure}
\includegraphics[width=0.8\textheight]{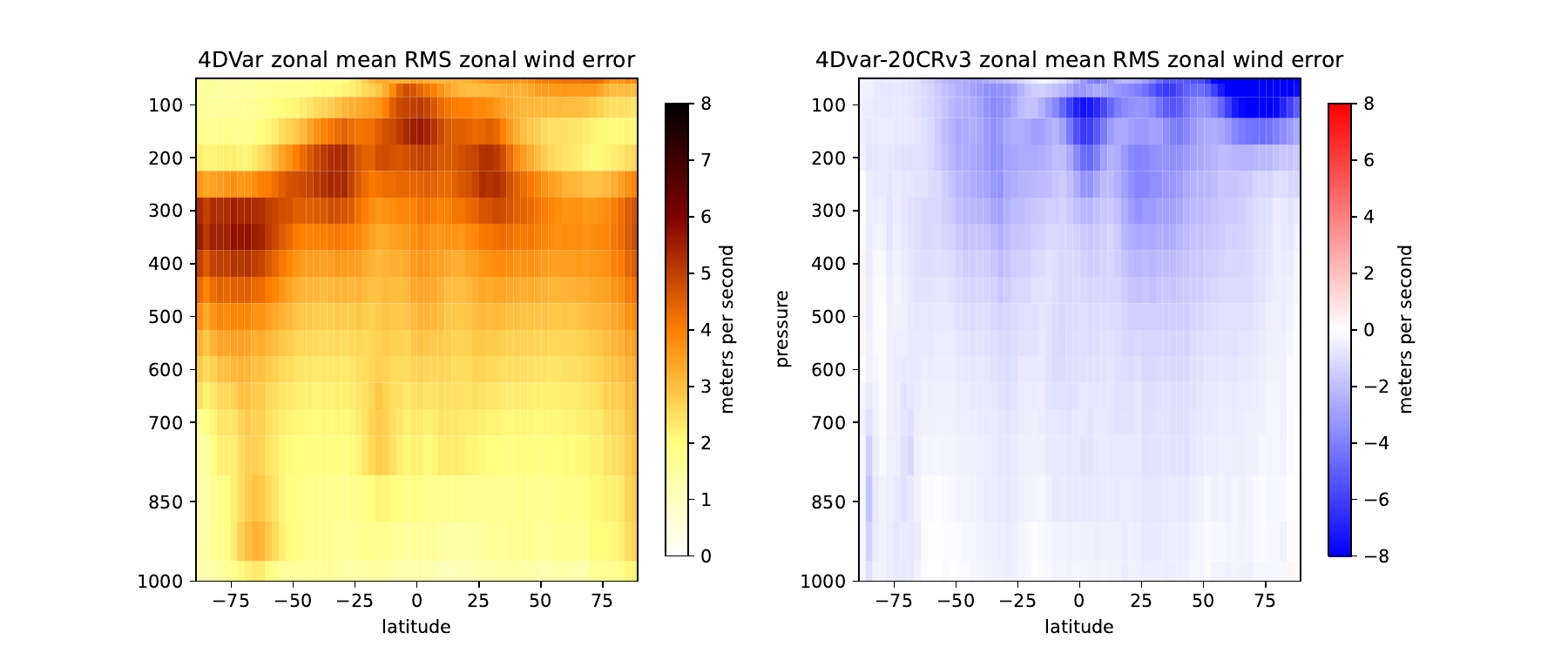}
\caption{Time mean zonally averaged zonal wind RMS error (relative to ERA5) for 4-day window 4DVar (left) and difference between 4-day 4DVar and 20CRv3 (left) for 362 analyses every 6-h starting January 1 2015.}
\label{zonalmeanuerr}
\end{figure}

\begin{figure}
  \centering
\includegraphics[width=1.0\textwidth, height=0.7\textheight, keepaspectratio]{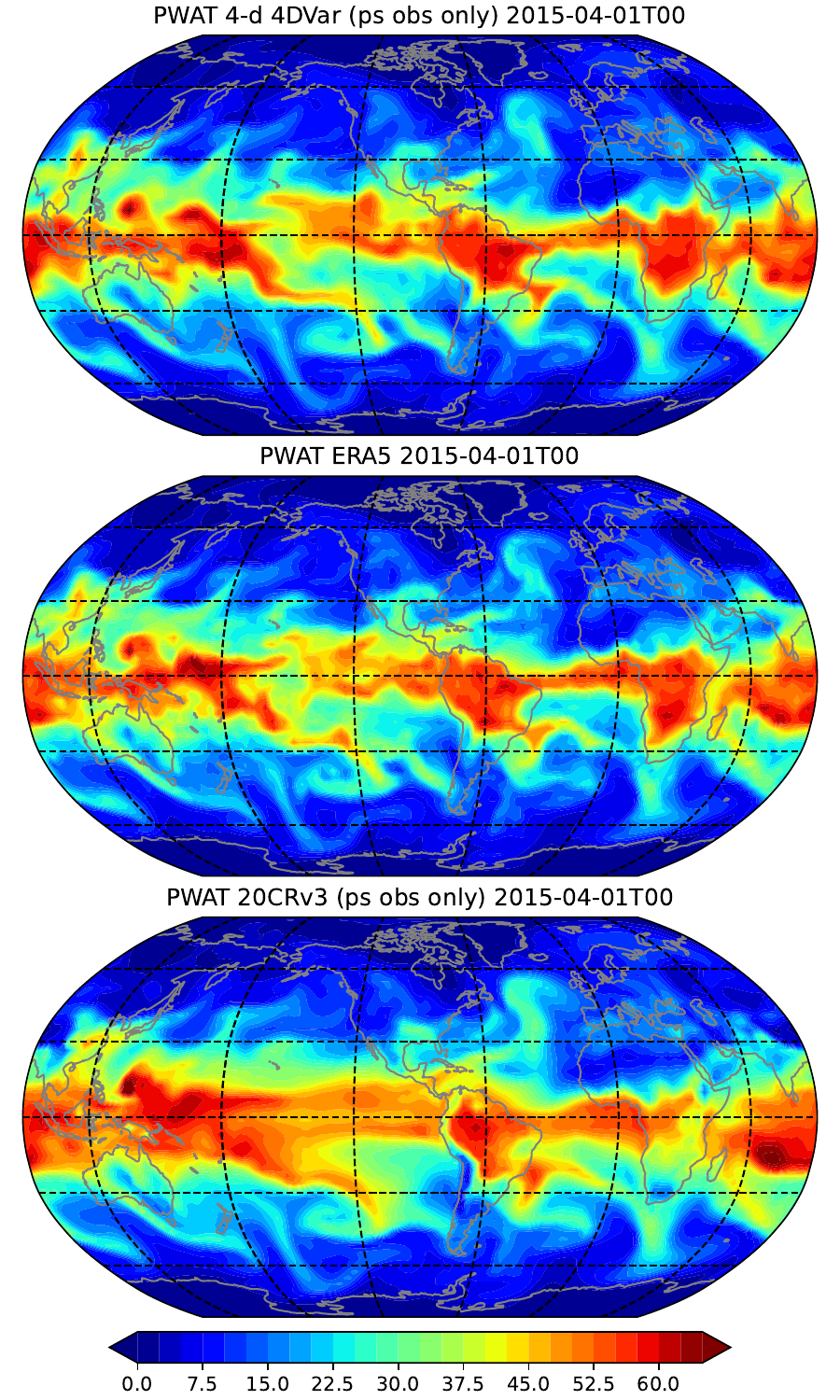}
\caption{4-day window 4DVar (top), ERA5 (middle), and 20CRv3 (bottom) analyses of precipitable water for 00UTC 1 April 2015 (after 360 analysis cycles).}
\label{pwat}
\end{figure}

Figure \ref{zonalmeanuerr} shows that the long window 4DVAR method can effectivity use surface pressure observations to consistently improve estimates of 3D atmospheric winds that are only indirectly related to surface pressure. The 4-day-window zonal-wind analyses are uniformly closer to ERA5 throughout the troposphere, and especially in the lower stratosphere.  The total integrated water vapor (precipitable water) field is even farther removed from surface pressure, and is not directly constrained by analysis optimization, since the gradients are explicitly set to zero in the minimization.  Figure \ref{pwat} compares the precipitable water analysis for the 4DVar 4-day window with ERA5 and 20CRv3 for 1 April 2015 at 00UTC, after 360 data assimilation cycles.  Although there are notable differences between the the 4-day 4DVar and ERA5 analyses, especially at smaller scales, most of the major synoptic-scale features present in the ERA5 are evident in the 4-day 4DVar analysis. Compared to 20CRv3, water vapor variability in the tropics is much better captured in the 4-day 4DVar analysis.  This suggests that the water vapor field is strongly constrained by the dynamics of the forecast model and the requirement that the trajectory remain close to the surface pressure observations in the analysis window.

\section{Conclusions \label{sec:conclusions}}

Strong-constraint long-window 4DVar using NeuralGCM and surface pressure observations is able to produce atmospheric state estimates that are significantly more accurate than the best sparse-input reanalysis currently available (20CRv3). With a window length of 4 days, it is possible to create 500hPa geopotential height analyses that are within about 10 m (in an RMS sense averaged over the Northern Hemisphere) of the current state-of-the-art full-input reanalysis (ERA5). The long window approach is made possible by recent advances in computing that were originally developed for training of artificial intelligence models. Specifically,  automatic differentiation of the NeuralGCM model based on the JAX framework, use of nonlinear gradient decent algorithms like AdamW that were originally developed to train neural networks, and computational efficiency of GPU-native models that avoid the need for explicitly linearized 4DVAR inner loops. Taken together, these advances, along with the neglect of the background and model-error terms in the loss function, allow the data assimilation code to be very concise and run efficiently on GPUs. 

 \citet{FisherAuvinen2012} show that the operational ECMWF analysis system has effectively no memory of observations and background states more than three days old. Therefore, a 4-day window should be long enough for the background term in the loss function to become irrelevant, except for the state estimates at the very beginning of the window.  Our experiments, which neglect the background term and instead use a simple tendency penalty to limit the impact of rapidly decaying modes at the start of the window, bear this out. We find that the state estimates become less accurate when the window length exceeds 4 days, which we suspect could have two contributions. One is from overfitting errors at the end of the window relative to early in the window \cite{PiresEtAl1996,vonich2026}, and the other is model error. The impact of neglecting model error in the loss function is yet to be determined. We hypothesize that the use of a weak-constraint formulation would improve the results, especially for longer windows.

\section*{Acknowledgments}
Discussions with Dr. Laura Slivinksi (NOAA/PSL) and Dr. Dan Holdaway (NOAA/NWS/OMD) are acknowledged. GJH acknowledges grant support by awards to the University of Washington from the National Science Foundation (award 2501400) and Heising-Simons Foundation (award 2023-4715). The statements, findings, conclusions, and recommendations are those of the authors and do not necessarily reflect the views of NOAA or the U.S. Department of Commerce.

\clearpage

\bibliographystyle{apalike}
\bibliography{refs}

\clearpage
\section*{Supporting Information}

\setcounter{figure}{0}
\renewcommand{\thefigure}{S\arabic{figure}}
\renewcommand{\theHfigure}{S\arabic{figure}}

\begin{figure}[htbp]
\centering
\includegraphics[width=0.95\textwidth]{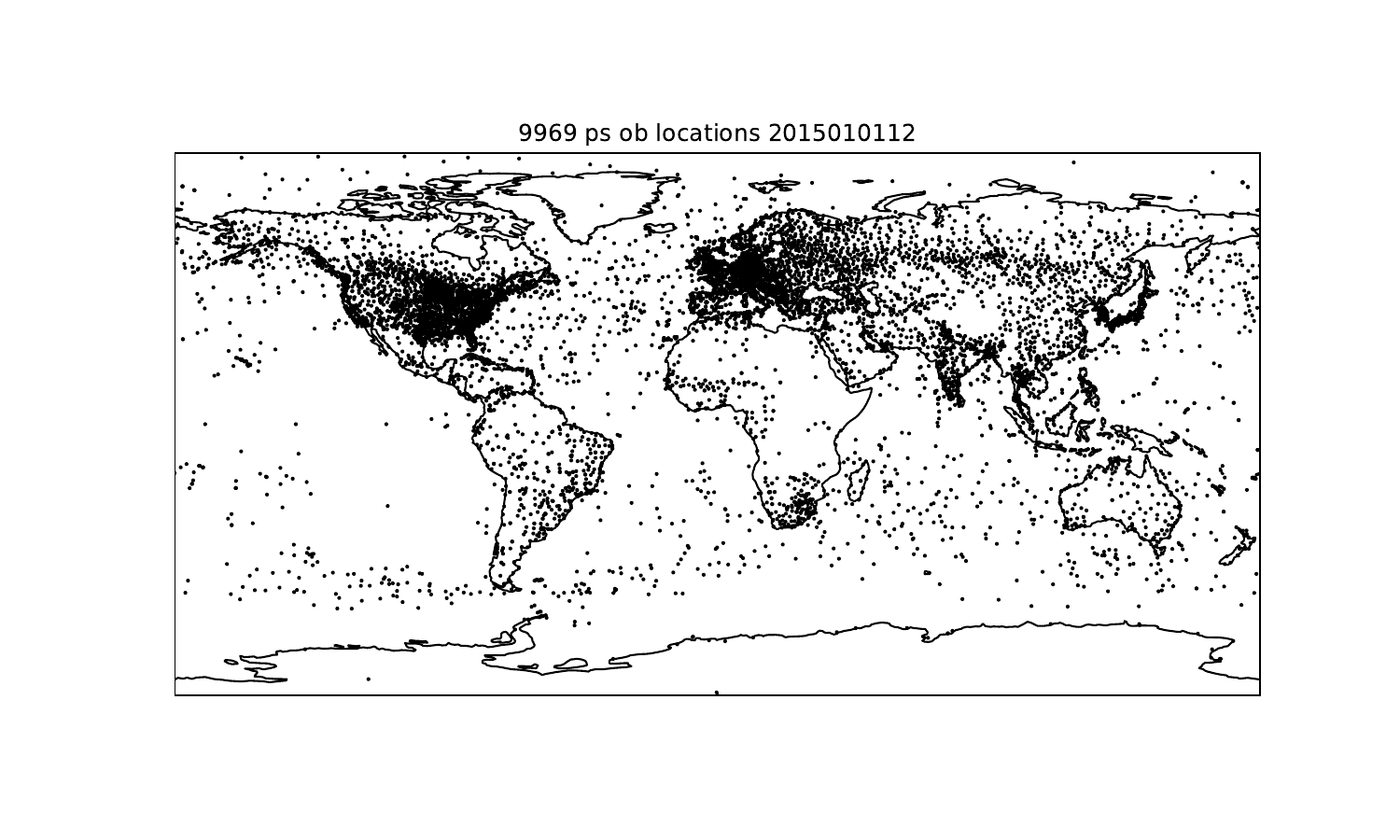}
\caption{Distribution of surface pressure observations on 1 January 2015 at 12UTC.}
\label{fig:s1}
\end{figure}

\begin{figure}[p]
\centering
\includegraphics[width=0.95\textwidth]{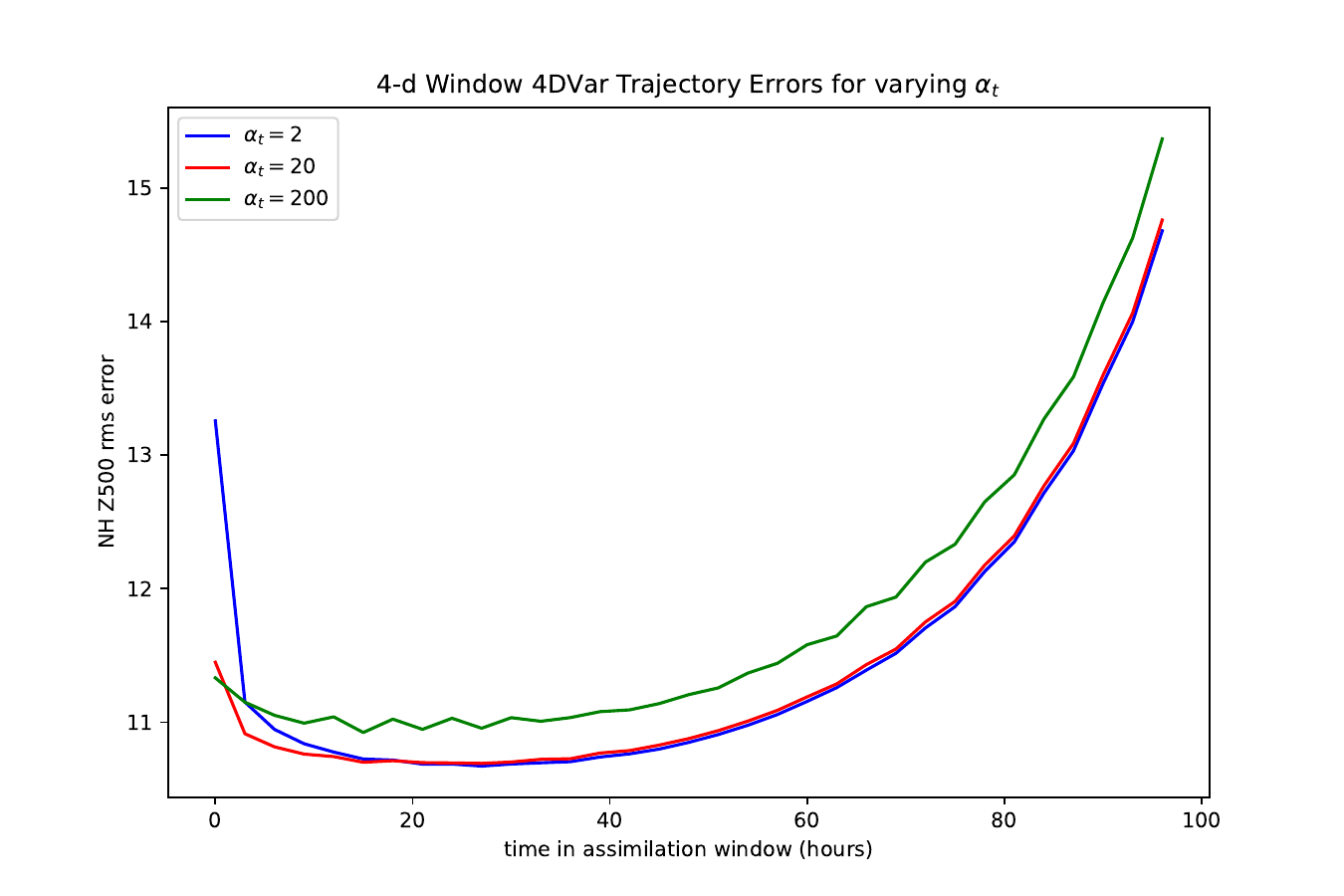}
\caption{Time averaged 500 hPa geopotential height (Z500) root mean squared error (relative to ERA5) over the assimilation window for 4-day window 4DVar experiments with varying coefficients for the amplitude of the tendency term in the loss ($\alpha_t$). Note that for $\alpha_t=2$ and 20 the Z500 error decreases rapidly over the first few hours of the window, asymptoting to the same value after about 18 hours. For $\alpha_t=2$ the mean RMS difference between the analysis and the observations at the first time in the window (0.96 hPa) is less than for $\alpha_t=20$ (1.01 hPa), even though the error in Z500 is larger. This reflects the impact of the rapidly decaying structures in the optimized initial condition that brings the state closer to the observations, but farther away from unobserved variables. For $\alpha_t=200$, the rapid decrease of Z500 error is eliminated, but the error is larger across the entire window, since less weight is given to the observation term in the loss during the optimization.}
\label{fig:s2}
\end{figure}

\begin{figure}[p]
\centering
\includegraphics[width=0.95\textwidth]{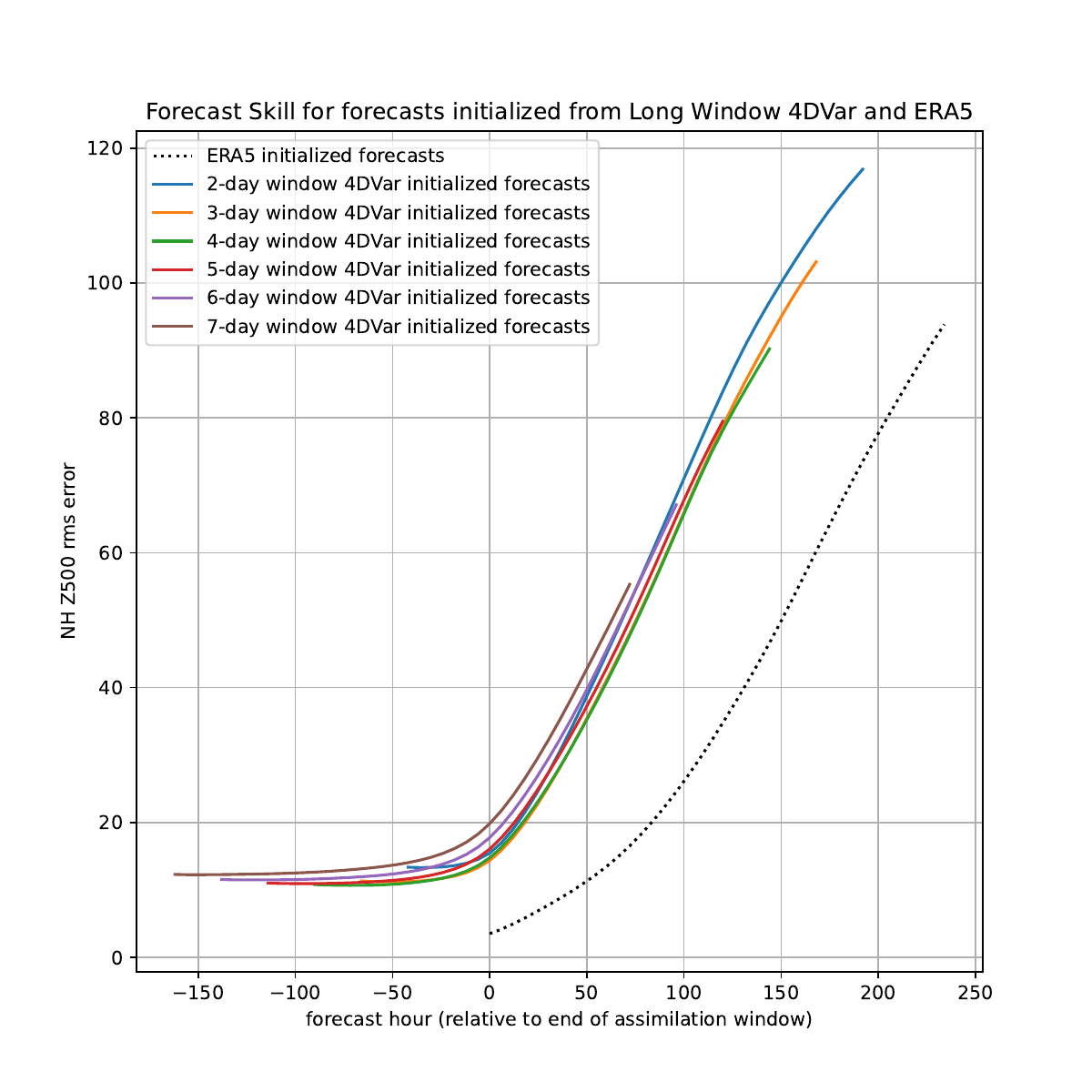}
\caption{500 hPa geopotential height forecast errors (in meters) averaged over the Northern Hemisphere (m) as a function of lead time for NeuralGCM initialized from ERA5 (black dotted line) and the optimal state estimates for the long window 4DVar window experiments for various window lengths (colored solid lines).  Forecast hour is defined relative to the end of the assimilation window (for ERA5 it was assumed that the analyses were at the center of the 12-h assimilation window).}
\label{fig:s3}
\end{figure}

\clearpage

\end{document}